\documentclass[aps,pra,reprint,superscriptaddress,longbibliography,nofootinbib,raggedbottom]{revtex4-2}

\usepackage{amsmath,amssymb,bm}
\usepackage{graphicx}
\usepackage{booktabs}
\usepackage{siunitx}
\DeclareSIUnit{\gauss}{G}
\usepackage{xcolor}
\usepackage{placeins}
\usepackage[colorlinks=true,allcolors=blue,hypertexnames=false]{hyperref}

\makeatletter
\newcommand{\PRAauthornote}[1]{%
  \@AF@join{\PRAauthornotemark{#1}}}
\newcommand{\PRAauthornotepair}[2]{%
  \@AF@join{\PRAauthornotemark{#1}\PRAauthornotemark{#2}}}
\newcommand{\PRAauthornotemark}[1]{\frontmatter@footnote{#1}}
\newcommand{\PRAauthornoteoutput}[1]{#1}
\def\doauthor#1#2#3{%
  \ignorespaces#1\unskip\@listcomma
  \begingroup
    #3%
  \@if@empty{#2}{\endgroup{}{}}{%
    \endgroup{\comma@space}{}\PRAauthornoteoutput{#2}}%
  \space\@listand
}
\makeatother

\begin{document}

\title{Stochastic Liouville--transport theory of light--atom interaction noise in thermal atomic vapors}
\author{Shaoxin Yuan}
\PRAauthornote{These authors contributed equally to this work.}
\author{Bin Wu}
\PRAauthornote{These authors contributed equally to this work.}
\author{Mingyong Jing}
\PRAauthornotepair
  {These authors contributed equally to this work.}
  {Corresponding author: \href{mailto:jmy@sxu.edu.cn}{jmy@sxu.edu.cn}}
\author{Chaoyang Hu}
\author{Yan Peng}
\author{Tingting Li}
\author{Xingya Li}
\author{Wenguang Yang}
\author{Junyao Xie}
\author{Zongkai Liu}
\author{Hao Zhang}
\author{Linjie Zhang}
\author{Liantuan Xiao}
\author{Suotang Jia}
\affiliation{State Key Laboratory of Quantum Optics Technologies and Devices, Institute of Laser Spectroscopy, Shanxi University, Taiyuan, Shanxi 030006, China}
\affiliation{Collaborative Innovation Center of Extreme Optics, Shanxi University, Taiyuan, Shanxi 030006, China}
\date{\today}

\begin{abstract}
Atom--light interaction noise can limit thermal-vapor sensing. Existing theories often treat internal-state dynamics, finite-mode atomic motion, and stochastic renewal separately, obscuring their coupled contributions to measured noise. We develop a general stochastic Liouville--transport theory, tested against polarization-resolved resonant Cs D$_2$ spectra. Joint experiment--theory analysis identifies atom--light noise below approximately \SI{100}{\kilo\hertz} as transit-dominated. Ballistic motion through the finite Gaussian mode modulates both the coupling-weighted effective atom number and trajectory-dependent Rabi coupling, producing predominantly common-mode noise. Boundary renewal introduces atoms with independently sampled ground-state sublevels, generating differential population fluctuations with opposite effects on the circular channels. Under an applied longitudinal magnetic field, experiment and theory show the same qualitative nonmonotonic change in common-mode suppression, supporting Zeeman redistribution of the channel responses. The framework can analyze noise in other thermal-atom sensors, including Rydberg-atom electric-field measurements.
\end{abstract}

\maketitle

\section{Introduction}

Thermal atomic vapors support compact, continuous quantum precision measurements without laser cooling or particle trapping~\cite{Kitching2018,Schlossberger2024}. Their applications include vapor-cell clocks~\cite{Knappe2004,Micalizio2021}, subfemtotesla optically pumped magnetometers~\cite{Kominis2003}, and Rydberg electrometers. Rydberg-vapor sensing has progressed from EIT--Autler--Townes field measurement~\cite{Sedlacek2012} to phase-resolved mixing~\cite{Simons2019Phase} and atomic superheterodyne detection of field amplitude, phase, and frequency~\cite{Jing2020Superhet}. It now supports SI-traceable~\cite{Holloway2014}, vector~\cite{Elgee2024}, and multichannel wideband measurements~\cite{Prajapati2025}, while atom-number scaling and receiver-noise mechanisms have become central to further sensitivity improvement~\cite{Zhang2023Scaling,Wang2023Noise}.

Thermal motion couples internal-state evolution to external dynamics: velocities set Doppler detunings, trajectories modulate finite-mode coupling, and atoms continuously leave and enter the interaction region while populations and coherences evolve. Existing theories treat selected parts of this coupled problem through phase-space kinetic or finite-beam models for ensemble-averaged spectra~\cite{Rautian1991,Firstenberg2008,Xiao2008}, linearized steady-state propagation theories for optical fluctuations~\cite{Lezama2008}, analytical correlation functions or stochastic fields for specified absorption and spin observables~\cite{Mitsui2013,Aoki2016,Lucivero2017}, and system-level receiver-noise models~\cite{Wang2023Noise,Zhang2023Scaling,Tang2025Noise}. These approaches obtain tractable solutions for their target regimes by selecting the physical processes and observables most relevant to each problem. In realistic thermal-atom measurements, however, multilevel evolution, Doppler shifts, finite-mode transit, quantum jumps, and continuous atomic exchange act together in the same measured noise spectrum. A more general and accurate microscopic theory that treats these coupled processes jointly is therefore urgently needed.

Here we formulate a stochastic Liouville--transport theory in terms of an operator-valued atomic phase-space density, coupling multilevel open-system dynamics to Doppler-resolved ballistic transport and stochastic atom renewal in a finite optical mode. We apply it to the complete 48-state Cs D$_2$ hyperfine--Zeeman manifold. With one common scale factor, the calculated polarization-resolved PSDs reproduce the measured low-frequency spectral dependence and the separation between the single-arm and balanced-difference spectra, while the calculated longitudinal-field dependence agrees qualitatively with experiment. The formulation can be extended to optical--Rydberg manifolds with microwave coupling for microscopic noise-source attribution in Rydberg-atom electric-field measurements.

\section{Liouville--transport theory}
\label{sec:theory}

The Liouville--transport formulation is independent of the atomic species and optical-readout configuration. Section~\ref{sec:procedures} specializes the model to Cs D$_2$.

\begin{figure*}[t!]
\centering
\includegraphics[width=0.95\textwidth]{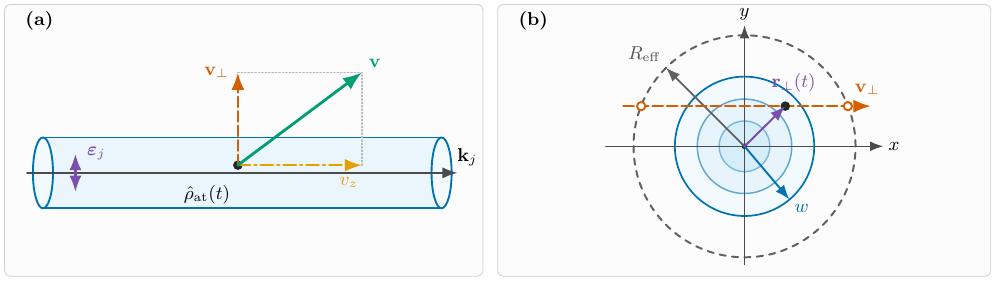}
\caption{Single-atom Liouville--transport dynamics. (a) Longitudinal excitation for a representative drive channel $j$ and decomposition of the atomic velocity. (b) Transverse ballistic trajectory through the circular Gaussian mode and open renewal boundary.}
\label{fig:transport_geometry}
\end{figure*}

For a dilute thermal vapor, the ensemble is described by the operator-valued phase-space density $\hat\varrho(\mathbf r_\perp,\mathbf v,t)$, which resolves the internal state as a function of transverse position and velocity. Its diagonal and off-diagonal elements are the corresponding population and coherence densities, while $\operatorname{Tr}\hat\varrho$ is the total atomic phase-space density. Writing $\mathbf r_\perp=(x,y)$, $\mathbf v=(v_x,v_y,v_z)$, $\mathbf v_\perp=(v_x,v_y)$, and $\nabla_\perp=(\partial_x,\partial_y)$, we adopt the collisionless, recoil-free ballistic limit, in which the coupled internal--external dynamics is
\begin{widetext}
\begin{equation}
\frac{\partial \hat\varrho}{\partial t}
=
-\mathbf v_\perp\cdot\nabla_\perp\hat\varrho
+\mathcal L_{\mathrm{int}}[\hat\varrho]
+\hat\Sigma_{\mathrm{in}}
-
\hat\Sigma_{\mathrm{out}} .
\label{eq:liouville_transport}
\end{equation}
\end{widetext}
The transport term $-\mathbf v_\perp\cdot\nabla_\perp\hat\varrho$ describes ballistic motion across the optical mode, and $\mathcal L_{\mathrm{int}}$ governs the internal state. The operator-valued rates $\hat\Sigma_{\mathrm{in}}$ and $\hat\Sigma_{\mathrm{out}}$ describe the injection of atoms into and their removal from the interaction region. Figure~\ref{fig:transport_geometry} shows the corresponding single-atom picture. Along a trajectory, $\hat\rho_{\mathrm{at}}(t)$ is the normalized internal-state density operator; its conditional ensemble average at fixed $\mathbf r_\perp$ and $\mathbf v$ is $\hat\varrho/\operatorname{Tr}\hat\varrho$. Each drive channel contributes to $\mathcal L_{\mathrm{int}}$ through its Doppler shift and position-dependent coupling.

\subsection{Single-atom internal-state evolution}
\label{subsec:internal_single_atom}

Along an individual ballistic trajectory between boundary events, the internal state obeys a single-atom open-system equation. To construct this equation, let $\mathcal K$ denote the retained fine-structure levels. A level $\ell\equiv(n_\ell,L_\ell,J_\ell)$ defines an electronic--nuclear subspace $\mathcal H_\ell$ spanned by the product states
\begin{equation}
\begin{aligned}
\lvert \ell;\alpha\rangle
&\equiv
\lvert n_\ell L_\ell J_\ell m_{J,\alpha}\rangle\\[-0.3ex]
&\quad\otimes
\lvert I,m_{I,\alpha}\rangle,\\
\ell&\in\mathcal K,
\qquad
\alpha=(m_{J,\alpha},m_{I,\alpha}) .
\end{aligned}
\label{eq:generic_basis}
\end{equation}
Here $n_\ell$ is the principal quantum number, $L_\ell$ and $J_\ell$ are the electronic orbital and total angular momenta, and $I$ is the nuclear spin. The projections range over $m_{J,\alpha}=-J_\ell,\ldots,J_\ell$ and $m_{I,\alpha}=-I,\ldots,I$. Thus $\mathcal H_\ell$ contains $(2J_\ell+1)(2I+1)$ product states. The retained internal space and the projector onto $\mathcal H_\ell$ are
\(
\mathcal H=\bigoplus_{\ell\in\mathcal K}\mathcal H_\ell
\)
and
\(
\hat\Pi_\ell=\sum_\alpha\lvert\ell;\alpha\rangle\langle\ell;\alpha\rvert,
\)
respectively. The basis of $\mathcal H$ is the corresponding union of the product bases.

Within $\mathcal H_\ell$, the scalar $E_\ell^{(0)}$ is the electronic energy common to the entire fine-structure level. The hyperfine operator $\hat H_{\mathrm{hfs}}$ couples the electronic total angular momentum $\hat{\mathbf J}$ to the nuclear spin $\hat{\mathbf I}$. Within each fine-structure level, its magnetic-dipole and electric-quadrupole couplings are parameterized by the conventional hyperfine constants $A_\ell$ and $B_\ell$, respectively. For a static magnetic field $\mathbf B_0$, the Zeeman operator $\hat H_Z(\mathbf B_0)=-(\hat{\boldsymbol\mu}_J+\hat{\boldsymbol\mu}_I)\cdot\mathbf B_0$ couples the electronic and nuclear magnetic moments to the field. In the field range considered, the hyperfine and Zeeman energy scales are small compared with the fine-structure separations, so the hyperfine--Zeeman Hamiltonian is treated separately within each $\mathcal H_\ell$, with
\[
\hat H_{\mathrm{hZ}}^{(\ell)}
\equiv
\hat\Pi_\ell
\left[\hat H_{\mathrm{hfs}}+\hat H_Z(\mathbf B_0)\right]
\hat\Pi_\ell .
\]
Let $\mathsf H_{\mathrm{hZ}}^{(\ell)}$ be the product-basis matrix of this operator. It is diagonalized as
\begin{equation}
\mathsf U_\ell^\dagger
\mathsf H_{\mathrm{hZ}}^{(\ell)}
\mathsf U_\ell
=
\mathsf E_{\mathrm{hZ}}^{(\ell)},
\qquad \ell\in\mathcal K .
\label{eq:static_diagonalization}
\end{equation}
The columns of $\mathsf U_\ell$ contain the product-basis coefficients of the hyperfine--Zeeman eigenstates, while the corresponding diagonal entries of $\mathsf E_{\mathrm{hZ}}^{(\ell)}$ give their energy shifts relative to $E_\ell^{(0)}$.

With the hyperfine--Zeeman eigenbasis established, let $j=1,\ldots,N_{\mathrm d}$ label the electric-dipole drive channels. Channel $j$ couples the ordered pair of levels $a_j\rightarrow b_j$. Its field has angular frequency $\omega_j$, wave vector $\mathbf k_j$, positive carrier amplitude $\mathcal E_{j,0}$, unit polarization vector $\boldsymbol{\varepsilon}_j$, and a slowly varying complex mode envelope $u_j$. We use the real-field convention
\[
\begin{aligned}
\mathbf E_j(\mathbf r,t)
&=
\frac{\mathcal E_{j,0}}{2}
u_j(\mathbf r_\perp)\boldsymbol{\varepsilon}_j
\\[-0.3ex]
&\quad\times
e^{\mathrm i(\mathbf k_j\cdot\mathbf r-\omega_jt)}
+\mathrm{c.c.},\\
\mathcal I_{j,0}&=\frac{1}{2}c\varepsilon_0\mathcal E_{j,0}^2 .
\end{aligned}
\]
Here $|u_j(\mathbf0)|=1$, so $\mathcal I_{j,0}$ is the on-axis cycle-averaged intensity in vacuum. Relative to the chosen quantization axis, we introduce the spherical unit vectors
\[
\mathbf e_0=\mathbf e_z,
\qquad
\mathbf e_{\pm1}
=
\mp\frac{\mathbf e_x\pm\mathrm i\mathbf e_y}{\sqrt2}.
\]
The field polarization is decomposed in this basis as
\[
\boldsymbol{\varepsilon}_j
=
\sum_{q=-1}^{+1}\epsilon_{j,q}\mathbf e_q,
\qquad
\sum_{q=-1}^{+1}|\epsilon_{j,q}|^2=1.
\]
Here $\epsilon_{j,q}$ is the normalized complex amplitude of the $q$-polarized field component, so its carrier amplitude is $\mathcal E_{j,0}\epsilon_{j,q}$. Thus $q=-1,0,+1$ labels $\sigma^-$, $\pi$, and $\sigma^+$, respectively. Any common carrier phase is included in $u_j$.

Independently of the applied field, the spherical components of the atomic electric-dipole operator are
\[
\hat d_q
\equiv
\mathbf e_q\cdot\hat{\mathbf d}.
\]
For channel $j$, let $\alpha$ and $\beta$ label product states in $\mathcal H_{a_j}$ and $\mathcal H_{b_j}$, respectively. The field-independent electric-dipole matrix element is
\begin{equation}
\begin{aligned}
d_{\beta\alpha}^{(j,q)}
&\equiv
\langle b_j;\beta\rvert
\hat d_q
\lvert a_j;\alpha\rangle\\
&=
\delta_{m_{I,\beta},m_{I,\alpha}}
(-1)^{J_{b_j}-m_{J,\beta}}
\begin{pmatrix}
J_{b_j} & 1 & J_{a_j}\\
-m_{J,\beta} & q & m_{J,\alpha}
\end{pmatrix}\\[-0.3ex]
&\quad\times
\langle b_j\Vert\hat{\mathbf d}\Vert a_j\rangle .
\end{aligned}
\label{eq:transition_dipole_element}
\end{equation}
The double bars denote the $J$-reduced dipole matrix element between the two fine-structure levels, and the parentheses denote a Wigner $3j$ symbol. The Kronecker delta enforces $m_{I,\beta}=m_{I,\alpha}$. The $3j$ symbol enforces $m_{J,\beta}=m_{J,\alpha}+q$ and sets the relative amplitude and phase of each allowed transition.

Combining the field-component amplitude $\mathcal E_{j,0}\epsilon_{j,q}$ with the corresponding dipole matrix in Eq.~\eqref{eq:transition_dipole_element} and then transforming from the product bases with $\mathsf U_{a_j}$ and $\mathsf U_{b_j}$ gives the $q$-resolved and total Rabi matrices for channel $j$:
\begin{equation}
\begin{aligned}
\boldsymbol{\Omega}_{j,q}
&=
-\frac{\mathcal E_{j,0}\epsilon_{j,q}}{\hbar}
\mathsf U_{b_j}^\dagger
\left[d_{\beta\alpha}^{(j,q)}\right]_{\beta,\alpha}
\mathsf U_{a_j},\\
\boldsymbol{\Omega}_{j}
&=
\sum_{q=-1}^{+1}
\boldsymbol{\Omega}_{j,q} .
\end{aligned}
\label{eq:dipole_rabi}
\end{equation}
The corresponding operators on the complete internal space are
\[
\begin{aligned}
\hat\Omega_{j,q}
&\equiv
-\frac{\mathcal E_{j,0}\epsilon_{j,q}}{\hbar}
\hat\Pi_{b_j}\hat d_q\hat\Pi_{a_j},
\\
\hat\Omega_j
&=
\sum_{q=-1}^{+1}\hat\Omega_{j,q} .
\end{aligned}
\]
Here $\boldsymbol{\Omega}_{j,q}$ is the matrix of $\hat\Omega_{j,q}$ between the source and target eigenbases. The operator $\hat\Omega_j$ maps $\mathcal H_{a_j}$ to $\mathcal H_{b_j}$, and $\hat\Omega_j^\dagger$ gives the reverse term in the electric-dipole interaction Hamiltonian.

To separate the optical carrier oscillations from the slower internal dynamics, we transform to a multifrequency rotating frame. For each subspace $\mathcal H_\ell$, we choose a scalar reference energy $E_{\ell,\mathrm{ref}}$ and define its shift from the fine-structure energy by $\delta E_{\ell,\mathrm{ref}}\equiv E_{\ell,\mathrm{ref}}-E_\ell^{(0)}$.
For an atom with velocity $\mathbf v$, the rotating angular frequency assigned to $\mathcal H_\ell$ is $E_{\ell,\mathrm{ref}}/\hbar+\Delta_\ell(\mathbf v)$, where $\Delta_\ell(\mathbf v)$ is the detuning assigned to that subspace. For a connected acyclic level-coupling graph, we choose one root subspace $\mathcal H_{\ell_0}$, set $\Delta_{\ell_0}\equiv0$, and determine the remaining detunings from
\begin{equation}
\Delta_{b_j}(\mathbf v)-\Delta_{a_j}(\mathbf v)
=
\omega_j-\mathbf k_j\cdot\mathbf v
-\frac{E_{b_j,\mathrm{ref}}-E_{a_j,\mathrm{ref}}}{\hbar}.
\label{eq:multidrive_detunings}
\end{equation}
The right-hand side is the angular-frequency detuning of drive $j$ in the atomic frame relative to the transition frequency defined by $E_{a_j,\mathrm{ref}}$ and $E_{b_j,\mathrm{ref}}$. With the mode envelope evaluated along the trajectory as $u_j(t)\equiv u_j[\mathbf r_\perp(t)]$, the rotating-wave approximation gives
\begin{equation}
\begin{aligned}
\hat H(t;\mathbf v)
&=
\sum_{\ell\in\mathcal K}
\left\{
\left[
\hat H_{\mathrm{hZ}}^{(\ell)}
-\delta E_{\ell,\mathrm{ref}}\hat\Pi_\ell
\right]
-\hbar\Delta_\ell(\mathbf v)\hat\Pi_\ell
\right\}\\
&\quad+
\frac{\hbar}{2}
\sum_{j=1}^{N_{\mathrm d}}
\left[
u_j(t)\hat\Omega_j
+u_j^*(t)\hat\Omega_j^\dagger
\right].
\end{aligned}
\label{eq:hamiltonian_full}
\end{equation}
In the first sum of Eq.~\eqref{eq:hamiltonian_full}, $\hat H_{\mathrm{hZ}}^{(\ell)}-\delta E_{\ell,\mathrm{ref}}\hat\Pi_\ell$ gives the hyperfine--Zeeman energies relative to $E_{\ell,\mathrm{ref}}$, while $-\hbar\Delta_\ell(\mathbf v)\hat\Pi_\ell$ gives the rotating-frame detuning of $\mathcal H_\ell$. The second sum contains the retained electric-dipole couplings. Dissipation is described by the Lindblad master equation
\begin{equation}
\begin{aligned}
\dot{\hat\rho}_{\mathrm{at}}
&=
\mathcal L_{\mathrm{int}}[\hat\rho_{\mathrm{at}}]\\
&=
-\frac{\mathrm{i}}{\hbar}[\hat H(t;\mathbf v),\hat\rho_{\mathrm{at}}]
+\sum_\xi \mathcal D_{\hat L_\xi}[\hat\rho_{\mathrm{at}}],\\
\mathcal D_{\hat L}[\hat\rho]
&=
\hat L\hat\rho\hat L^\dagger
-\frac{1}{2}\{\hat L^\dagger\hat L,\hat\rho\}.
\end{aligned}
\label{eq:lindblad_master}
\end{equation}
Here $\hat L_\xi$ is the Lindblad operator for incoherent channel $\xi$, $\mathcal D_{\hat L}$ is the corresponding Lindblad superoperator, and $[\cdot,\cdot]$ and $\{\cdot,\cdot\}$ denote the commutator and anticommutator, respectively. Radiative decay, population transfer, and dephasing among the retained states are represented by the corresponding $\hat L_\xi$. All Lindblad operators are represented in the same hyperfine--Zeeman eigenbasis and transformed consistently to the rotating frame used in Eq.~\eqref{eq:hamiltonian_full}. The spontaneous-emission operators used for Cs D$_2$ are specified in Sec.~\ref{subsec:model_protocol}. Equation~\eqref{eq:lindblad_master}, with the Hamiltonian in Eq.~\eqref{eq:hamiltonian_full}, explicitly defines the internal generator $\mathcal L_{\mathrm{int}}$ in Eq.~\eqref{eq:liouville_transport}; the ballistic-transport and boundary-renewal terms are specified next.

\subsection{Ballistic transport and boundary renewal}
\label{subsec:external_single_atom}

During its residence in $\mathcal R$, each atom samples the transverse optical profile along its ballistic path [Fig.~\ref{fig:transport_geometry}]. With the normalization $|u_j(\mathbf 0)|=1$ introduced above, the local intensity of field $j$ for an arbitrary centered profile is related to its mode envelope by
\begin{equation}
\frac{\mathcal I_j(\mathbf r_\perp)}{\mathcal I_{j,0}}
=
\left|u_j(\mathbf r_\perp)\right|^2 .
\label{eq:mode_definition}
\end{equation}
Here $\mathcal I_j(\mathbf r_\perp)$ is the transverse intensity. Through $u_j(t)\hat\Omega_j$ in Eq.~\eqref{eq:hamiltonian_full}, ballistic motion converts the spatial profile sampled by the atom into a time-dependent coupling. The Gaussian probe mode used below is
\begin{equation}
u_{\mathrm G}(\mathbf r_\perp)
=
\exp\!\left(
-\frac{x^2}{w_x^2}
-\frac{y^2}{w_y^2}
\right),
\label{eq:mode_profiles}
\end{equation}
where $w_x$ and $w_y$ are the $1/e^2$ intensity radii along $x$ and $y$, respectively. In this transverse-mode description, variation along $z$ is neglected.

The ballistic motion and circular renewal region are defined by
\begin{equation}
\frac{d\mathbf r_\perp}{dt}=\mathbf v_\perp,
\qquad
\frac{d\mathbf v}{dt}=0,
\qquad
\mathcal R=
\{\mathbf r_\perp:|\mathbf r_\perp|<R_{\mathrm{eff}}\} .
\label{eq:external_motion}
\end{equation}
Here $R_{\mathrm{eff}}$ is the renewal-boundary radius. A trajectory ends when the atom reaches $|\mathbf r_\perp|=R_{\mathrm{eff}}$ while moving outward; its evolved internal state then leaves $\mathcal R$, implementing $\hat\Sigma_{\mathrm{out}}$ in Eq.~\eqref{eq:liouville_transport}.

To complete the open-boundary model, we specify the total injection rate and the position, velocity, and internal-state statistics of atoms injected into $\mathcal R$. The surrounding vapor follows the Maxwell--Boltzmann (MB) velocity distribution
\begin{equation}
\phi_{\mathrm{MB}}(\mathbf v)
=
\left(\frac{m}{2\pi k_BT}\right)^{3/2}
\exp\!\left[-\frac{m|\mathbf v|^2}{2k_BT}\right] .
\label{eq:maxwell}
\end{equation}
Here $m$ is the atomic mass, $T$ is the vapor temperature, $k_B$ is the Boltzmann constant, and $\sigma_v=\sqrt{k_BT/m}$ is the one-dimensional velocity standard deviation.

For a spatially uniform reservoir surrounding the circular region, injection positions are uniformly distributed in azimuth, while the injected velocities are weighted by their inward flux. At the boundary, let $\mathbf n$ be the outward transverse unit normal and define the inward normal speed $v_n=-\mathbf v_\perp\cdot\mathbf n>0$. Atoms crossing inward during $dt$ originate from a layer of thickness $v_n\,dt$, so the differential inward flux is proportional to $v_n\phi_{\mathrm{MB}}(\mathbf v)d^3v$. Conditional on an inward crossing, the normalized velocity distribution is the flux-weighted Maxwellian~\cite{Rautian1991,Flekkoy2005}
\begin{equation}
\phi_{\mathrm{in}}(\mathbf v\mid\mathbf n)
=
\frac{\sqrt{2\pi}}{\sigma_v}
v_n\phi_{\mathrm{MB}}(\mathbf v),
\qquad
v_n>0.
\label{eq:flux_distribution}
\end{equation}
Equivalently, $p_{\mathrm{in}}(v_n)$ is the normalized marginal probability density of the inward normal speed:
\begin{equation}
p_{\mathrm{in}}(v_n)
=
\frac{v_n}{\sigma_v^2}
\exp\!\left(-\frac{v_n^2}{2\sigma_v^2}\right),
\qquad
v_n>0,
\label{eq:normal_flux_distribution}
\end{equation}
while the transverse velocity component tangent to the boundary and the longitudinal component $v_z$ retain their Maxwellian distributions.

The internal state of the injected atoms is described by the density operator $\hat\rho_{\mathrm{in}}$, which is set by the surrounding reservoir or any preceding state-preparation process. The density operator used for the Cs D$_2$ calculation is given in Eq.~\eqref{eq:rho_in}. The total injection rate and the distributions of injection position and inward velocity, together with $\hat\rho_{\mathrm{in}}$, determine the incoming source term $\hat\Sigma_{\mathrm{in}}$. These injection statistics and the outward-crossing rule complete the boundary renewal in Eq.~\eqref{eq:liouville_transport}.

\subsection{Trajectory representation and \texorpdfstring{PSD estimation}{PSD estimation}}
\label{subsec:stochastic_observables}

For each trajectory, the injection position and inward velocity are sampled from the distributions above, and the initial pure internal state is drawn from the prescribed ensemble whose mean density operator is $\hat\rho_{\mathrm{in}}$. The atom then follows Eq.~\eqref{eq:external_motion} while its internal state evolves through a Monte Carlo wave-function (MCWF) unraveling~\cite{Dalibard1992,Molmer1993} of Eq.~\eqref{eq:lindblad_master}; an outward crossing ends the trajectory. Ensemble averaging over the sampled injection positions, velocities, initial pure states, and quantum-jump histories reproduces the evolution in Eq.~\eqref{eq:liouville_transport}.

Within each trajectory, the MCWF representation assigns the atom a normalized pure state $\lvert\psi(t)\rangle$. Between quantum jumps, the state evolves under the non-Hermitian Hamiltonian. The continuous-time evolution and infinitesimal jump probabilities are
\begin{equation}
\begin{aligned}
\hat H_{\mathrm{eff}}(t;\mathbf v)
&=
\hat H(t;\mathbf v)
-\frac{\mathrm{i}\hbar}{2}
\sum_\xi \hat L_\xi^\dagger \hat L_\xi,\\
\frac{d}{dt}\lvert\widetilde\psi(t)\rangle
&=
-\frac{\mathrm{i}}{\hbar}
\hat H_{\mathrm{eff}}(t;\mathbf v)\lvert\widetilde\psi(t)\rangle,\\
dp_\xi
&=
\langle\psi(t)\rvert
\hat L_\xi^\dagger \hat L_\xi
\lvert\psi(t)\rangle\,dt .
\end{aligned}
\label{eq:mcwf_update}
\end{equation}
Here $\lvert\widetilde\psi\rangle$ is the unnormalized no-jump state, whereas $\lvert\psi\rangle$ is the normalized state used to evaluate the instantaneous channel rates. The total jump probability in $dt$ is $\sum_\xi dp_\xi$. If a jump occurs, channel $\xi$ is selected with probability proportional to $dp_\xi$, and the state becomes
\begin{equation}
\lvert\psi\rangle
\longrightarrow
\frac{\hat L_\xi\lvert\psi\rangle}
{\lVert \hat L_\xi\lvert\psi\rangle\rVert}.
\label{eq:mcwf_jump}
\end{equation}

The readout trace $s(t)$ is formed by summing the mode-weighted responses of all atoms currently within $\mathcal R$. The polarization-resolved traces used for the Cs D$_2$ calculation are defined in Eqs.~\eqref{eq:mode_projected_coherence} and \eqref{eq:absorptive_normalization}. For a stationary real trace with mean $\overline s=\langle s(t)\rangle$, the one-sided PSD at $f>0$ is defined directly from the finite-time Fourier transform of its fluctuation:
\begin{equation}
S_s(f)
=
\lim_{\mathcal T\rightarrow\infty}
\frac{2}{\mathcal T}
\left\langle
\left|
\int_0^{\mathcal T}
\left[s(t)-\overline s\right]
e^{-\mathrm{i}2\pi ft}\,dt
\right|^2
\right\rangle.
\label{eq:psd_definition}
\end{equation}
Here $\langle\cdot\rangle$ denotes averaging over the stochastic trajectory realizations, $\mathcal T$ is the record duration, and $f$ is the positive Fourier frequency. The factor of $2$ specifies the one-sided convention.

\section{Experimental methods and Cs D\texorpdfstring{$_2$}{2} implementation}
\label{sec:procedures}

The general theory becomes a quantitative prediction once the optical geometry, atomic structure, renewal statistics, and measured observable are specified. We now specify these ingredients for the Cs D$_2$ experiment.

\begin{figure*}[t]
\centering
\includegraphics[width=0.98\textwidth]{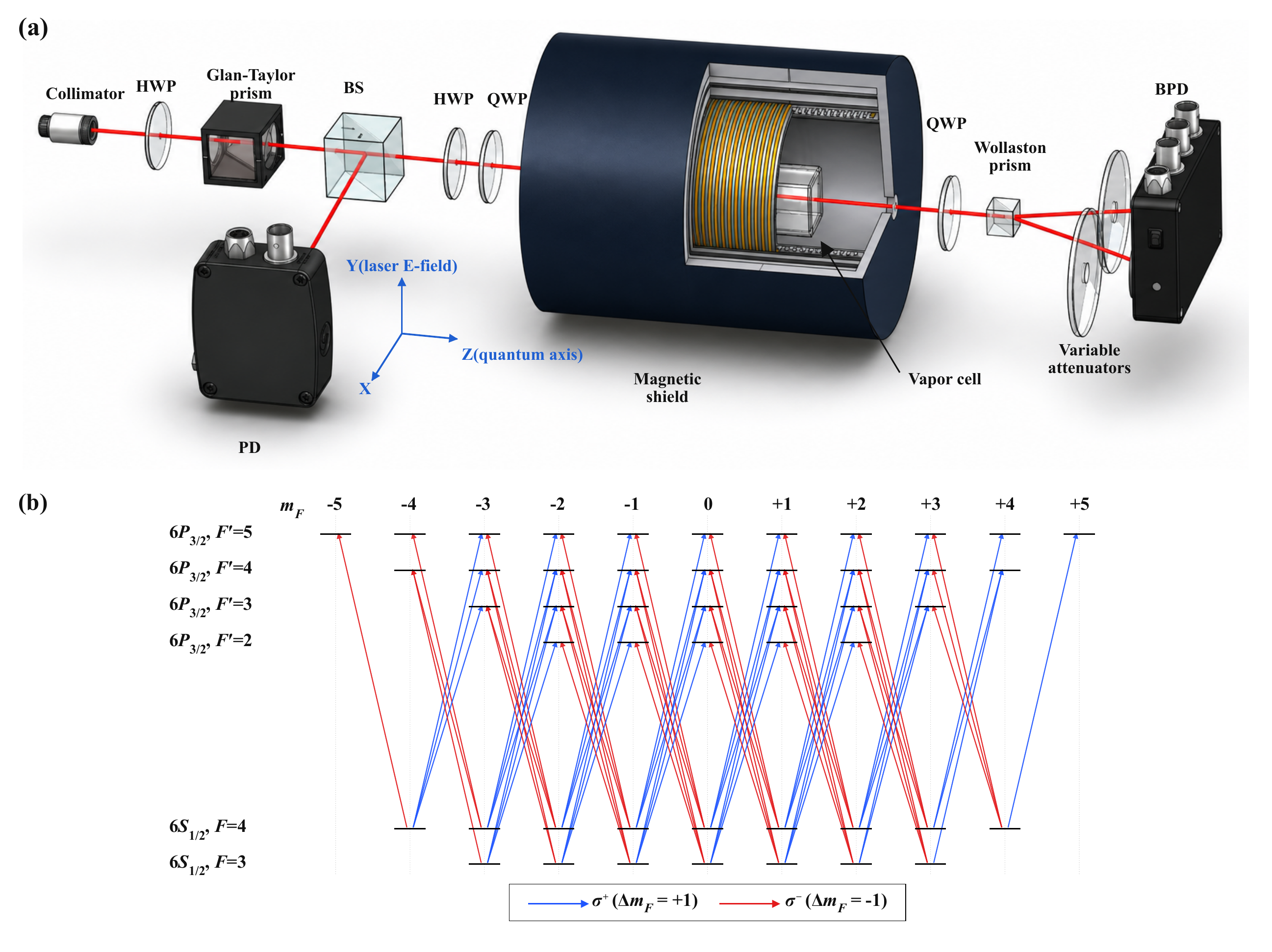}
\caption{Experimental apparatus and Cs D$_2$ coupling scheme. (a) Optical layout for single-arm and balanced detection of the circular components. HWP, half-wave plate; QWP, quarter-wave plate; BS, beam splitter; PD, photodetector; BPD, balanced photodetector. (b) Hyperfine manifolds retained in the 48-state model. Blue and red arrows indicate the allowed $\Delta m_F=+1$ and $-1$ couplings, respectively.}
\label{fig:setup}
\end{figure*}

\subsection{Experimental apparatus and acquisition}
\label{subsec:experimental_setup}

We record fluctuations in the two circular components of the transmitted probe both near zero field within the magnetic shield and at several longitudinal static-field values [Fig.~\ref{fig:setup}(a)]. The vertically polarized, cavity-stabilized probe is tuned to the Cs D$_2$ $F=4\rightarrow F'=5$ resonance. It has a wavelength of $\lambda_{\mathrm p}=\SI{852.347}{\nano\meter}$, a total power of $P_{\mathrm p}\simeq\SI{580}{\micro\watt}$, and a circular Gaussian transverse profile with a $1/e^2$ radius of $w=\SI{0.75}{\milli\meter}$.

The cell has outer dimensions \SI{8.4}{\centi\meter}$\times$\SI{2.4}{\centi\meter}$\times$\SI{2.4}{\centi\meter}, with the first dimension along the probe axis. The two \SI{0.2}{\centi\meter}-thick windows give an optical path length of \SI{8}{\centi\meter} through the Cs vapor. A solenoid provides the longitudinal static magnetic field, $\mathbf B_0\parallel\mathbf k$ (Faraday geometry). During the field scan, the probe carrier frequency is held fixed, and only the field magnitude $B_0$ is varied. For the near-zero-field data set, the residual field inside the magnetic shield is below \SI{2}{\nano\tesla}, so the Zeeman shifts are negligible relative to the optical linewidth.

After the cell, a QWP and Wollaston prism separate the transmitted probe into its $\sigma^+$ and $\sigma^-$ components with an extinction ratio of approximately $10^6{:}1$. Let $P_\pm(t)$ denote the optical powers incident on the two balanced-detector inputs after attenuation, and define $\delta P_\pm(t)=P_\pm(t)-\overline P_\pm$. Variable attenuators set $\overline P_+=\overline P_-\equiv\overline P_{\mathrm{det}}=\SI{10}{\micro\watt}$. The normalized single-arm traces are
\[
S_\pm(t)
=
\frac{\delta P_\pm(t)}{\overline P_{\mathrm{det}}}.
\]
With matched calibrated responsivities and electronic gains, the normalized balanced trace is
\begin{equation}
S_3(t)
\equiv
S_+(t)-S_-(t).
\label{eq:balanced_readout}
\end{equation}
For an $S_+$ or $S_-$ measurement, the unused detector input is blocked while the active input remains at $\overline P_{\mathrm{det}}$. For an $S_3$ measurement, both inputs remain illuminated and balanced. The subtraction in Eq.~\eqref{eq:balanced_readout} suppresses common-mode fluctuations.

Each of the three channels is recorded using a spectrum analyzer with a resolution bandwidth of \SI{200}{\hertz}. For each channel, a far-detuned reference is acquired by shifting the probe several gigahertz from resonance while keeping the detector power, gain, and acquisition settings identical to those of the corresponding resonant measurement. These references provide the optical-detection baselines, and the photodetector (PD) noise floor is measured separately. Before comparison with the calculation, all resonant, far-detuned, and PD-floor PSDs undergo the low-frequency correction described in Appendix~\ref{app:experimental_preprocessing}.

\subsection{Theoretical model for Cs D\texorpdfstring{$_2$}{2}}
\label{subsec:model_protocol}

For the Cs D$_2$ calculation, $\mathcal K=\{g,e\}$, with $g\equiv6S_{1/2}$ and $e\equiv6P_{3/2}$. A single probe couples $g$ to $e$. The ground-state subspace $\mathcal H_g$ contains seven $F=3$ and nine $F=4$ Zeeman states, while the excited-state subspace $\mathcal H_e$ contains 32 states; the full internal space $\mathcal H=\mathcal H_g\oplus\mathcal H_e$ is therefore 48-dimensional [Fig.~\ref{fig:setup}(b)]. Because there is only one drive, we henceforth suppress the label $j$ and write $\boldsymbol{\Omega}=\sum_q\boldsymbol{\Omega}_q$ and $\hat\Omega=\sum_q\hat\Omega_q$.

For each applied $\mathbf B_0$, the hyperfine--Zeeman Hamiltonian is diagonalized separately in $\mathcal H_g$ and $\mathcal H_e$ using Eq.~\eqref{eq:static_diagonalization}. The resulting eigenstates and energies define the internal Hamiltonian and the basis in which the dipole matrices below are represented. We approximate the magnetically shielded near-zero-field condition by $\mathbf B_0=0$; the eigenstates can then be labeled by $F$ and $m_F$: $\lvert g;F,m_F\rangle$ with $F=3,4$, and $\lvert e;F',m_F'\rangle$ with $F'=2,3,4,5$. The corresponding hyperfine-level projectors are
\[
\begin{aligned}
\hat\Pi_{gF}
&\equiv
\sum_{m_F=-F}^{F}
\lvert g;F,m_F\rangle\langle g;F,m_F\rvert,
\\
\hat\Pi_g
&=\sum_{F=3}^{4}\hat\Pi_{gF},\\
\hat\Pi_{eF'}
&\equiv
\sum_{m_F'=-F'}^{F'}
\lvert e;F',m_F'\rangle\langle e;F',m_F'\rvert,
\\
\hat\Pi_e
&=\sum_{F'=2}^{5}\hat\Pi_{eF'} .
\end{aligned}
\]

At zero field, let $F_g\equiv F$ and $F_e\equiv F'$. For $\ell=g,e$, define the usual Casimir factor

\[
\begin{aligned}
K_{\ell F_\ell}
&=F_\ell(F_\ell+1)-I(I+1)\\[-0.2ex]
&\quad-J_\ell(J_\ell+1).
\end{aligned}
\]

\begin{samepage}
\noindent
The corresponding ground- and excited-state hyperfine energies satisfy
\[
E_{gF}-E_g^{(0)}=\frac{A_g}{2}K_{gF},
\]
\[
\begin{aligned}
E_{eF'}-E_e^{(0)}
&=\frac{A_e}{2}K_{eF'}
+\frac{B_e}{2I(2I-1)J_e(2J_e-1)}\\[-0.2ex]
&\quad\times\Bigl[
\tfrac{3}{4}K_{eF'}(K_{eF'}+1)\\[-0.2ex]
&\hspace{5.0em}{}-I(I+1)J_e(J_e+1)
\Bigr].
\end{aligned}
\]
\end{samepage}

Thus $A_g$ and $A_e$ are the magnetic-dipole hyperfine constants of $6S_{1/2}$ and $6P_{3/2}$, respectively, whereas $B_e$ is the electric-quadrupole hyperfine constant of $6P_{3/2}$. Expressed in frequency units, the values used here are $A_g/h\simeq\SI{2298.158}{\mega\hertz}$, $A_e/h\simeq\SI{50.288}{\mega\hertz}$, and $B_e/h\simeq-\SI{0.493}{\mega\hertz}$, with $B_g=0$. For a longitudinal field $\mathbf B_0=B_0\mathbf e_z$, the Zeeman term within $\mathcal H_\ell$ is
\[
\hat H_Z^{(\ell)}
=
\frac{\mu_BB_0}{\hbar}
\left(
g_{J,\ell}\hat J_z
+g_I^{\mathrm{eff}}\hat I_z
\right).
\]
The calculation uses $g_{J,g}\simeq2.002$, $g_{J,e}\simeq1.334$, the effective nuclear factor $g_I^{\mathrm{eff}}\simeq-3.989\times10^{-4}$, and $\mu_B/h\simeq\SI{13.996}{\giga\hertz\per\tesla}$~\cite{SteckCs2025}. Together, these hyperfine and Zeeman constants specify the matrices diagonalized in Eq.~\eqref{eq:static_diagonalization}; their zero-field eigenenergies are the $E_{gF}$ and $E_{eF'}$ used below.
Choosing $g$ as the root subspace and using $E_{g4}$ and $E_{e5}$ as the reference energies for $\mathcal H_g$ and $\mathcal H_e$, respectively, gives

\[
\begin{aligned}
\Delta_g
&\equiv0,\\
\Delta_e(\mathbf v)
&=\Delta-\mathbf k\!\cdot\!\mathbf v,\\
\Delta
&\equiv
\omega_{\mathrm{probe}}
-\frac{E_{e5}-E_{g4}}{\hbar}.
\end{aligned}
\]

Here $|\mathbf k|=2\pi/\lambda_{\mathrm p}$. Equation~\eqref{eq:hamiltonian_full} then reduces to
\begin{equation}
\begin{aligned}
\hat H_{\mathrm{D2}}(t;\mathbf v)
&=
\sum_{F=3}^{4}
(E_{gF}-E_{g4})\hat\Pi_{gF}
\\
&\quad+
\sum_{F'=2}^{5}
\bigl[
E_{eF'}-E_{e5}
-\hbar(\Delta-\mathbf k\!\cdot\!\mathbf v)
\bigr]
\hat\Pi_{eF'}
\\
&\quad+
\frac{\hbar}{2}
\left[
u_{\mathrm G}(t)\hat\Omega
+u_{\mathrm G}^*(t)\hat\Omega^\dagger
\right].
\end{aligned}
\label{eq:hamiltonian_cs_d2}
\end{equation}
For the near-zero-field resonant calculation, Eq.~\eqref{eq:hamiltonian_cs_d2} is evaluated with $\Delta=0$, while every other dipole-allowed hyperfine branch is retained at its physical detuning. Applied-field calculations use the same carrier frequency and the full Hamiltonian in Eq.~\eqref{eq:hamiltonian_full}, constructed in the field-dependent eigenbasis.

To specify the optical coupling in Eq.~\eqref{eq:hamiltonian_cs_d2}, we choose the probe propagation axis as the quantization axis, with the applied field parallel to it. The vertically polarized probe has $\epsilon_{+1}=\epsilon_{-1}=1/\sqrt2$ and $\epsilon_0=0$, so only the $q=\pm1$ components enter the driven coupling. For the circular Gaussian mode, $w_x=w_y=w$. The on-axis Rabi-frequency scale corresponding to the total field amplitude $\mathcal E_0$ is computed with the Alkali Rydberg Calculator (ARC)~\cite{Sibalic2017} from the experimental probe power and $1/e^2$ beam radius specified in Sec.~\ref{subsec:experimental_setup}; the products $\mathcal E_0\epsilon_q$ enter the Rabi matrices through Eq.~\eqref{eq:dipole_rabi}. We denote the resulting beam-center total-field Rabi-frequency scale by $\Omega_0$.

To complete the Lindblad dynamics, spontaneous emission from $6P_{3/2}$ to $6S_{1/2}$ is represented by jump operators with total population-decay rate $\Gamma_2$, where $\Gamma_2/(2\pi)=\SI{5.234}{\mega\hertz}$~\cite{SteckCs2025}. For $q=-1,0,+1$, let $\mathcal P_q$ denote the allowed decay paths associated with the corresponding spherical dipole component. Thus spontaneous emission includes the $q=0$ component as well as $q=\pm1$. For a path $p\in\mathcal P_q$, let $\zeta_p^{(q)}$, $\omega_p$, and $\hat A_p=\lvert g_p\rangle\langle e_p\rvert$ denote its dimensionless dipole amplitude, transition angular frequency, and lowering operator. For each excited eigenstate $\lvert e_a\rangle$, the amplitudes are normalized according to $\sum_{q=-1}^{+1}\sum_{p:e_p=e_a}|\zeta_p^{(q)}|^2=1$. Different emitted-polarization classes are treated as orthogonal, while paths within a given class are combined through the phenomenological Lorentzian frequency-overlap matrix
\begin{equation}
\begin{aligned}
(\mathsf K^{(q)})_{pp'}
&=
\frac{\Gamma_2\zeta_p^{(q)}\zeta_{p'}^{(q)*}}
{1+[(\omega_p-\omega_{p'})/\Gamma_2]^2},\\
\mathsf K^{(q)}\boldsymbol w_{\mu q}
&=\kappa_{\mu q}\boldsymbol w_{\mu q},
\qquad \kappa_{\mu q}>0,\\
\hat L_{\mu q}
&=\sqrt{\kappa_{\mu q}}
\sum_{p\in\mathcal P_q}
(\boldsymbol w_{\mu q})_p\hat A_p .
\end{aligned}
\label{eq:decay_kossakowski}
\end{equation}
Here $\mu=1,\ldots,\operatorname{rank}\mathsf K^{(q)}$ labels the nonzero decay eigenchannels obtained by diagonalizing $\mathsf K^{(q)}$ at fixed $q$. Accordingly, the spontaneous-emission contribution to the dissipator in Eq.~\eqref{eq:lindblad_master} is summed over all $q$ and $\mu$. The resulting operators satisfy $\sum_{q=-1}^{+1}\sum_\mu\hat L_{\mu q}^\dagger\hat L_{\mu q}=\Gamma_2\hat\Pi_e$. For each applied field, the states, amplitudes, and transition frequencies in Eq.~\eqref{eq:decay_kossakowski} are recomputed in the corresponding hyperfine--Zeeman eigenbasis.

\subsection{Numerical realization and simulated observables}
\label{subsec:numerical_protocol}

With the Cs D$_2$ Hamiltonian and decay channels specified, we express angular frequencies in units of $\Gamma_2$ and use the dimensionless time $\widetilde t=\Gamma_2t$; SI units are restored for plotting. The incoming reservoir is represented by the unpolarized ground-state mixture
\begin{equation}
\hat\rho_{\mathrm{in}}
=
\frac{\hat\Pi_g}{16}.
\label{eq:rho_in}
\end{equation}
Accordingly, each initial or newly injected MCWF pure state is sampled independently and uniformly from the 16 field-dependent eigenstates in $\mathcal H_g$.

Each calculation uses $M=100$ independent realizations of $N=64$ atom samples, all initially active. The GPU implementation is summarized in Appendix~\ref{app:computational_implementation}. We use $R_{\mathrm{eff}}=10w=\SI{7.5}{\milli\meter}$, $T=\SI{319.697}{\kelvin}$, and a Cs atomic mass of $m\simeq132.906~\mathrm{u}$~\cite{SteckCs2025}, giving the one-dimensional velocity width $\sigma_v\simeq\SI{141.421}{\meter\per\second}$. Initial transverse positions are sampled uniformly in area within $|\mathbf r_\perp|<R_{\mathrm{eff}}$, and initial velocities are sampled from Eq.~\eqref{eq:maxwell}.

During each realization, a sample is deactivated when its atom crosses the boundary outward. Poisson injection events refill inactive samples with new atoms initialized just inside the boundary at uniformly sampled azimuths. Their inward velocities are sampled from Eq.~\eqref{eq:flux_distribution}. The injection rate is chosen to maintain a mean active count of approximately 64 atom samples.

The classical trajectory and internal state are evolved jointly from $\widetilde t=0$ to $\widetilde t=10^5$ with a time step of $\delta\widetilde t=10^{-3}$. Between jumps, the unnormalized state is advanced under $\hat H_{\mathrm{eff}}$ using a split-step method. The diagonal internal-energy, detuning, and non-Hermitian decay terms are applied exponentially, while the optical interaction is integrated by a fourth-order Runge--Kutta step. Radiative jumps follow Eqs.~\eqref{eq:mcwf_update} and \eqref{eq:mcwf_jump}.

The propagated states are used to construct the simulated optical readout. For each circular component, let $i=1,\ldots,N$ label the atom samples, $\nu=1,\ldots,M$ label the realizations, and $\chi_i^{(\nu)}(t)\in\{0,1\}$ indicate whether sample $i$ contains an atom. Writing $\lvert\psi_i^{(\nu)}\rangle=\lvert\psi_{i,g}^{(\nu)}\rangle\oplus\lvert\psi_{i,e}^{(\nu)}\rangle$, we define the mode-projected, Rabi-weighted coherence and its absorptive quadrature as
\begin{equation}
\begin{aligned}
r_q^{(\nu)}(t)
&=
\frac{1}{N}
\sum_{i=1}^{N}
\chi_i^{(\nu)}(t)
u_{\mathrm G}[\mathbf r_{i\perp}^{(\nu)}(t)]
\\[-0.3ex]
&\hspace{2.8em}\times
\langle\psi_{i,e}^{(\nu)}(t)\rvert
\hat\Omega_q
\lvert\psi_{i,g}^{(\nu)}(t)\rangle,\\
Q_q^{(\nu)}(t)
&=
\operatorname{Im}\!\left[r_q^{(\nu)}(t)\right],
\qquad q=\pm1 ,
\end{aligned}
\label{eq:mode_projected_coherence}
\end{equation}
Here $r_q^{(\nu)}$ is the complex mode-projected coherence, and $Q_q^{(\nu)}$ is its real-valued absorptive quadrature; both have units of angular frequency.

The calculated traces are normalized using the late-time ensemble means of these quadratures:

\begin{equation}
\begin{aligned}
Q_{q,\mathrm{ref}}
&=
\left\langle
\frac{1}{M}\sum_{\nu=1}^{M}Q_q^{(\nu)}(t)
\right\rangle_{\mathrm{final}\ 20\%},\\
S_\pm^{(\nu)}(t)
&=
\frac{Q_{\pm1}^{(\nu)}(t)-Q_{\pm1,\mathrm{ref}}}
{Q_{\pm1,\mathrm{ref}}},\\
S_3^{(\nu)}(t)
&=
S_+^{(\nu)}(t)
-S_-^{(\nu)}(t).
\end{aligned}
\label{eq:absorptive_normalization}
\end{equation}

Equation~\eqref{eq:absorptive_normalization} defines one trace for each realization. A PSD is estimated separately for each $S_\pm^{(\nu)}$ and $S_3^{(\nu)}$ trace using Welch's method~\cite{Welch1967} with 95\% overlap between adjacent segments, and the resulting PSDs are averaged over the $M$ realizations. The difference $S_3^{(\nu)}$ is formed within each realization before its PSD is evaluated. All calculated PSDs contain only fluctuations generated by the atomic model.

\section{Results and discussion}
\label{sec:results}

\subsection{Near-zero-field comparison with experiment}
\label{subsec:validation}

\begin{figure*}[t]
\centering
\includegraphics[width=0.86\textwidth]{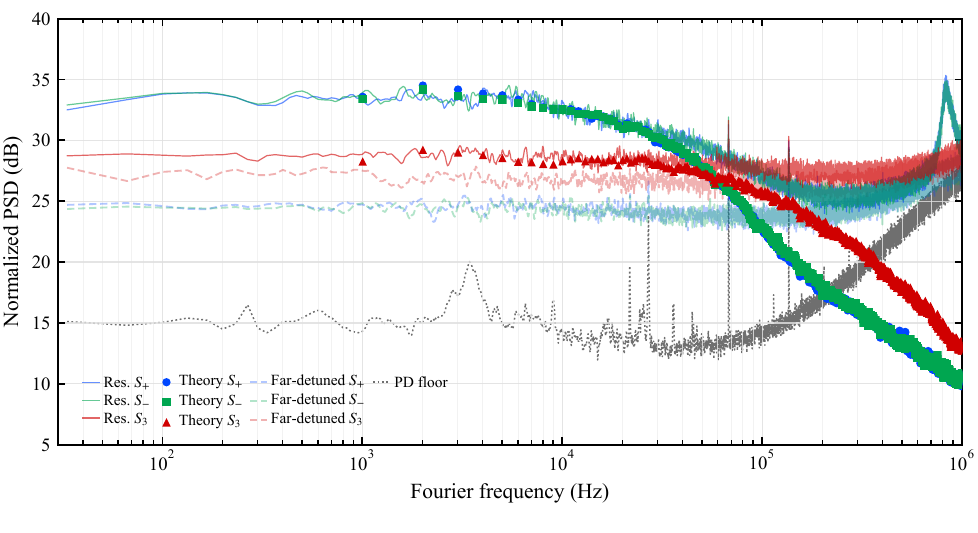}
\caption{Near-zero-field one-sided PSDs of $S_+$, $S_-$, and $S_3$. Solid and pale dashed curves show the corrected resonant measurements and far-detuned baselines, respectively; blue circles ($S_+$), green squares ($S_-$), and red triangles ($S_3$) show the atom-only calculations. The gray dotted curve is the photodetector noise floor.}
\label{fig:validation}
\end{figure*}

\begin{figure*}[t!]
\centering
\includegraphics[width=\textwidth]{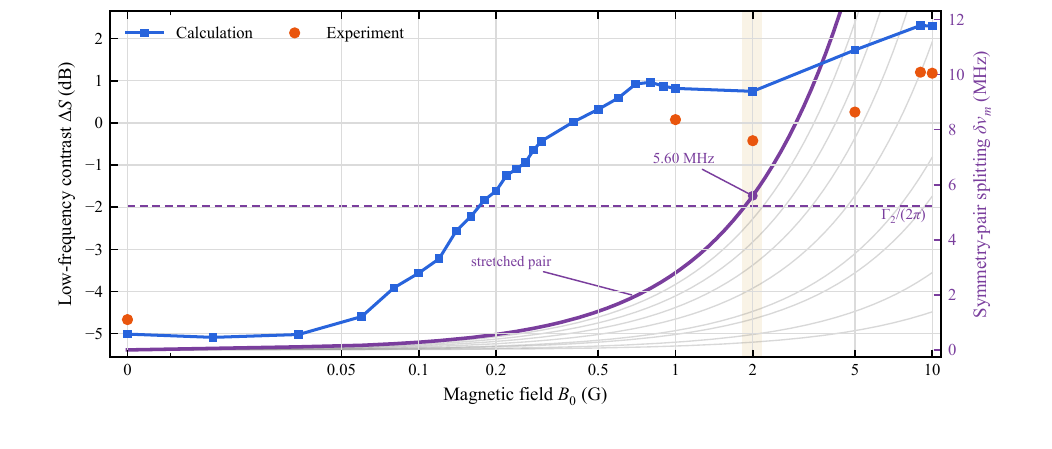}
\caption{Low-frequency contrast $\Delta S$ (left axis) and relative Zeeman splitting between symmetry-related $q=+1$ ($\sigma^+$) and $q=-1$ ($\sigma^-$) transitions (right axis). The purple curve denotes the stretched-state pair, and the dashed line marks $\Gamma_2/(2\pi)$.}
\label{fig:static_field_comparison}
\end{figure*}

A single dB offset is chosen so that the calculated and measured near-zero-field $S_+$ PSDs have the same mean over \SIrange{1}{5}{\kilo\hertz}; the same offset is applied to every channel and field setting. Near zero field, the resonant $S_+$ and $S_-$ PSDs are nearly equal, as expected from the $\sigma^+\leftrightarrow\sigma^-$ symmetry. From \SI{1}{\kilo\hertz} to about \SI{30}{\kilo\hertz}, both resonant single-arm PSDs decrease with increasing frequency and exhibit nearly identical spectral shapes.

Both single-arm PSDs exhibit a broad maximum near \SI{0.84}{\mega\hertz}. Between the low-frequency roll-off and this maximum, the resonant single-arm PSDs reach a shallow minimum at \SIrange{0.2}{0.3}{\mega\hertz}. At the minimum they lie only about \SI{1}{\decibel} above the far-detuned optical baselines, so the measured level is dominated by the optical-detection baseline, principally photon shot noise, with only a small resonant excess remaining. The proximity of this maximum to the frequency-lock servo bandwidth identifies it as residual probe-frequency noise shaped by the servo response. At zero field, the nearly degenerate transitions addressed by the two circular components have almost identical frequency-to-power responses, so this noise is predominantly common mode and is strongly suppressed in $S_3$.

The full transport calculation reproduces the low-frequency dependence of the normalized single-arm PSDs, including their shared roll-off near \SI{30}{\kilo\hertz}. For the Gaussian intensity weight $W(\mathbf r_\perp)\equiv|u_{\mathrm G}(\mathbf r_\perp)|^2=\exp(-2r_\perp^2/w^2)$, define the transit correlation $C_{\mathrm{tr}}(\tau)\equiv\langle W[\mathbf r_\perp(t)]W[\mathbf r_\perp(t+\tau)]\rangle$, where the average is over uniformly distributed transverse positions and Maxwellian transverse velocities; $C_{\mathrm{tr}}(0)$ is its zero-delay value. Ballistic averaging gives $C_{\mathrm{tr}}(\tau)/C_{\mathrm{tr}}(0)=[1+2(\sigma_v\tau/w)^2]^{-1}$, whose Fourier transform is proportional to $\exp(-\sqrt{2}\pi wf/\sigma_v)$. Its half-power frequency is
\begin{equation}
f_{\mathrm{tr},1/2}
=
\frac{(\ln 2)\sigma_v}{\sqrt{2}\pi w}
=\SI{29}{\kilo\hertz},
\label{eq:transit_scale_estimate}
\end{equation}
for $w=\SI{0.75}{\milli\meter}$ and $\sigma_v=\sqrt{k_BT/m}=\SI{141.4}{\meter\per\second}$. This value agrees with the roll-off near \SI{30}{\kilo\hertz} in both the full calculation and the measured PSDs. Using the same scaling for all channels, the calculation also reproduces the experimentally observed separation between $S_3$ and $S_\pm$ [Fig.~\ref{fig:validation}]. This agreement identifies the shared roll-off near \SI{30}{\kilo\hertz} as a predominantly common-mode finite-transit response whose spectral scale is set by ballistic motion through the nonuniform Gaussian mode. Open-boundary replacement maintains the stationary thermal ensemble, while the differential contribution driven by the incoming ground-state sublevels is considered below.

Balanced subtraction strongly suppresses the predominantly common-mode transit response and servo peak. At low frequencies, the resonant $S_3$ PSD remains above the directly measured far-detuned baseline. As the residual atomic contribution decreases, independent photon shot noise becomes dominant and the measured resonant $S_3$ PSD approaches the far-detuned baseline. Toward the upper end of the measured band, both the resonant and far-detuned $S_3$ PSDs follow the rise of the independently measured PD floor, showing that the high-frequency behavior is detector limited [Fig.~\ref{fig:validation}].

In the model, the residual differential component is seeded by incoming thermal atoms whose ground-state sublevels are statistically independent and distributed according to the unpolarized reservoir mixture. Successive boundary-renewal events therefore drive fluctuations in the population difference between positive- and negative-$m_F$ sublevels within the optical mode. Because these fluctuations change the two circular absorptions with opposite signs, they survive balanced subtraction. This mechanism gives an atom-only $S_3$ half-power frequency of \SI{95.3}{\kilo\hertz}. The transit-free calculation in Appendix~\ref{app:density_matrix_response} identifies a related dominant nonoscillatory local response at \SI{112.7}{\kilo\hertz}: 98.3\% of the complete 16-initial-state squared modal amplitude is the $F=4$ component proportional to $m_F$, while the total $F=3$ contribution is negligible. The agreement of these frequencies, together with the $F=4$, $m_F$-proportional modal structure, strongly supports continual injection of thermal atoms from the external reservoir as the dominant mechanism renewing the residual differential response.

\subsection{Response under applied static magnetic fields}
\label{subsec:static_field_comparison}

Having separated the near-zero-field common-mode response from the residual differential response, we next use a longitudinal field to control circular-channel symmetry. We define $\Delta S$ as the \SIrange{1}{5}{\kilo\hertz} mean $S_3$ PSD level minus the mean of the two single-arm levels in decibels; $\Delta S<0$ denotes common-mode suppression. At the fixed probe frequency, Zeeman shifts lift the degeneracy of symmetry-related $m_F\rightarrow m_F'$ pathways, causing the two circular channels to weight magnetic sublevels and velocity classes differently and reducing their shared response.

The calculation places most of the initial loss of suppression below \SI{1}{\gauss}, consistent with the difference between the near-zero-field and \SI{1}{\gauss} measurements. Over \SIrange{1}{10}{\gauss}, the measurements qualitatively test the calculated field dependence: both show a shallow decrease near \SI{2}{\gauss}, a rise through \SI{9}{\gauss}, and little further change at \SI{10}{\gauss} [Fig.~\ref{fig:static_field_comparison}]. Near \SI{2}{\gauss}, the shallow feature occurs as the representative stretched-state separation becomes comparable to $\Gamma_2/(2\pi)$. The stretched-state pair reaches a separation comparable to the linewidth near the observed feature, suggesting a crossover in spectral overlap. However, the feature is not determined by this pair alone: the other Zeeman branches have different shifts and line strengths, and the sum of the complete transition manifold determines its sign and small magnitude. The calculated values remain approximately \SIrange{0.7}{1.5}{\decibel} above the measurements; longitudinal optical propagation and experimental detection backgrounds omitted from the atom-only calculation are plausible contributors. The qualitative agreement supports the role of Zeeman-induced circular-channel asymmetry in governing common-mode suppression.

\FloatBarrier
\section{Conclusion}

We developed a general theory for optical readout in thermal atomic vapors that couples stochastic Liouville dynamics to thermal transport. Resonant Cs D$_2$ spectra provide its present experimental test. With one common scale factor, the theory reproduces the measured low-frequency spectral dependence and the separation between the single-arm and balanced-difference spectra. The comparison resolves a noise hierarchy: servo-associated laser-frequency noise is predominantly common mode near zero field and is suppressed in $S_3$, detector electronics set the high-frequency floor, and independent photon shot noise adds in $S_3$. Under the resonant-probe conditions studied here, the atom--light interaction contribution below approximately \SI{100}{\kilo\hertz} is dominated by transit-related noise. Near zero field, the common-mode finite-transit response arises as random trajectories through the Gaussian mode vary both the coupling-weighted effective atom number and the local Rabi coupling governing each atom's internal response. Boundary renewal instead seeds the residual differential response: incoming atoms with independently sampled ground-state sublevels generate population-imbalance fluctuations between positive- and negative-$m_F$ states that affect the circular absorptions with opposite signs and survive balanced subtraction. A longitudinal field changes the magnetic-sublevel and velocity-class weightings of the circular channels, and the qualitative agreement between the measured and calculated field dependences supports this channel-correlation picture.

As the next step, we will extend this framework to optical--Rydberg manifolds with microwave coupling to attribute microscopic noise and guide sensitivity improvements in Rydberg-atom electric-field measurements.

\section*{Acknowledgments}

National Key R\&D Program of China (2022YFA1404003); National Natural Science Foundation of China (12574318,62201481); Sanjin Talent Program of Shanxi Province (SJYC2024247, SJYC2025280).

\section*{Author Contributions}

M.J. and S.Y. conceived the experimental approach. S.Y. and B.W. performed the experiments and collected the data. M.J. developed the theoretical framework and program. M.J., S.Y. and B.W. carried out the theoretical simulations and calculations. M.J. and S.Y. analyzed the data and interpreted the results. M.J., L.Z., and H.Z. contributed to the experimental setup. M.J. and S.Y. wrote the manuscript. All authors contributed to discussions of the results and the manuscript and provided revisions of the manuscript.

\section*{Conflict of Interest}

The authors declare no competing interests.

\section*{Data Availability}

The data that support the findings of this study are available from the corresponding author upon reasonable request.

\appendix
\makeatletter
\@addtoreset{figure}{section}
\makeatother
\renewcommand{\thefigure}{\thesection\arabic{figure}}
\renewcommand{\theHfigure}{appendix.\thesection.\arabic{figure}}

\section{Low-frequency correction of experimental PSDs}
\label{app:experimental_preprocessing}

\begin{figure}[t!]
\centering
\includegraphics[width=\columnwidth]{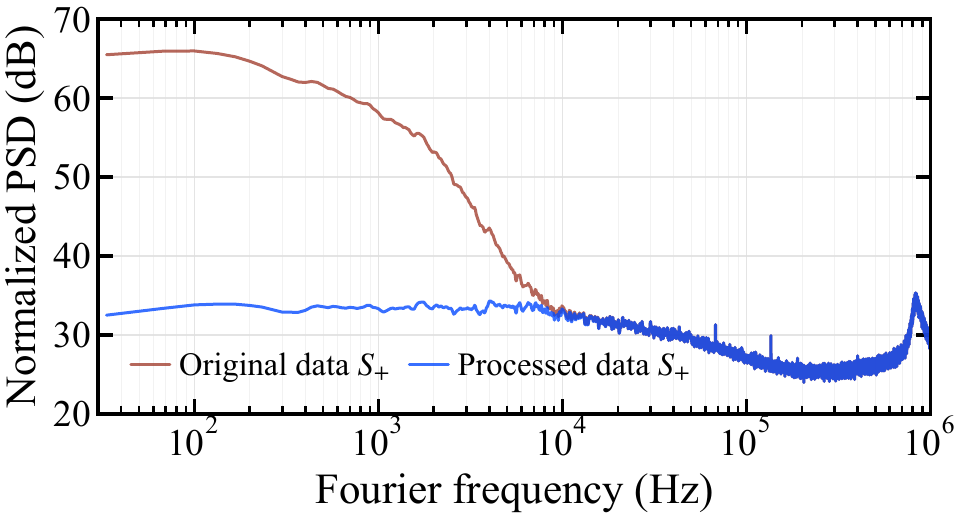}
\caption{Representative low-frequency correction of the near-zero-field resonant $S_+$ PSD. The red and blue curves show the original and processed spectra, respectively.}
\label{fig:experimental_preprocessing}
\end{figure}

All measured PSDs used in the comparison are corrected for a smooth low-frequency $1/f$-type background. The procedure is illustrated for the near-zero-field resonant $S_+$ PSD in Fig.~\ref{fig:experimental_preprocessing}; the same processed spectrum appears as the solid resonant $S_+$ measurement in Fig.~\ref{fig:validation}. Let $Y_{\mathrm{raw}}(f)$ denote the measured PSD expressed in decibels. For each trace, the median PSD near \SI{10}{\kilo\hertz} defines the reference level $Y_0$. With $Y_0$ fixed, we fit the low-frequency PSD samples directly by nonlinear least squares to

\begin{equation}
\begin{aligned}
Y_{\mathrm{fit}}(f)
&=
Y_0+\Delta Y_{\mathrm{LF}}(f),\\
\Delta Y_{\mathrm{LF}}(f)
&=
\sum_{r=1}^{2}
\frac{G_r}{1+\left(f/f_r\right)^{n_r}},
\end{aligned}
\label{eq:experimental_baseline}
\end{equation}
where $G_r$, $f_r$, and $n_r$ are, respectively, the fitted amplitude in decibels, corner frequency, and roll-off exponent for term $r$. The fitted low-frequency correction is subtracted from the measured PSD according to
\begin{equation}
Y_{\mathrm{corr}}(f)
=
Y_{\mathrm{raw}}(f)
-
\Delta Y_{\mathrm{LF}}(f).
\label{eq:experimental_baseline_correction}
\end{equation}
The remaining resonant, far-detuned, and PD-floor PSDs are corrected analogously, with $Y_0$ and the fit parameters determined independently for each data set.

\section{Local \texorpdfstring{$S_3$}{S3} response without spatial transit}
\label{app:density_matrix_response}

To isolate the internal $S_3$ response from spatial transit, we omit the ballistic-transport and boundary-renewal terms in Eq.~\eqref{eq:liouville_transport}. The local calculation evaluates the internal generator in Eq.~\eqref{eq:lindblad_master} at $\mathbf B_0=\mathbf0$, $\Delta=0$, $\mathbf r_\perp=\mathbf0$, and $\mathbf v=\mathbf0$, with the beam-center total-field Rabi scale $\Omega_0=5.436\Gamma_2$. We denote this time-independent generator by $\mathcal L_{\mathrm{loc}}$, so that the density matrix obeys
\begin{equation}
\frac{d\hat\rho_{\mathrm{at}}}{dt}
=
\mathcal L_{\mathrm{loc}}[\hat\rho_{\mathrm{at}}].
\label{eq:local_density_matrix_equation}
\end{equation}
We solve this equation separately for each ground-state Zeeman input:
\begin{equation}
\begin{gathered}
\hat\rho_{\mathrm{at}}(0)
=
\lvert g;F,m_F\rangle\langle g;F,m_F\rvert,\\
F=3,4,
\qquad
m_F=-F,\ldots,F.
\end{gathered}
\label{eq:local_initial_conditions}
\end{equation}
The $F=3$ and $F=4$ manifolds contain seven and nine initial states, respectively; each solution tracks the evolution of all populations and coherences from the specified initial sublevel.

Let $S_3(t\mid F,m_F)$ denote the difference-channel response obtained from the corresponding initial condition. The state $\hat\rho_{\mathrm{in}}=\hat\Pi_g/16$ in Eq.~\eqref{eq:rho_in} is an equally weighted mixture of these 16 initial states. Because the evolution in Eq.~\eqref{eq:local_density_matrix_equation} and the $S_3$ readout are linear, the response of this mixture is the arithmetic mean of the 16 state-resolved responses. We therefore define

\begin{equation}
\begin{aligned}
\delta S_{3,F,m_F}(t)
&=
S_3(t\mid F,m_F)\\
&\quad-
\frac{1}{16}
\sum_{\bar F=3}^{4}
\sum_{\bar m=-\bar F}^{\bar F}
S_3(t\mid \bar F,\bar m).
\end{aligned}
\label{eq:local_s3_density_matrix_response}
\end{equation}

The finite-dimensional matrix representation of $\mathcal L_{\mathrm{loc}}$ is diagonalized numerically, giving

\begin{equation}
\delta S_{3,F,m_F}(t)
=
\sum_j
a_{F,m_F}^{(j)}
\exp(\lambda_j t).
\label{eq:local_s3_modal_expansion}
\end{equation}

Here $j$ labels the distinct Liouvillian eigenvalues that contribute to the $S_3$ readout, and $a_{F,m_F}^{(j)}$ is the corresponding modal amplitude. A real negative $\lambda_j$ gives a nonoscillatory term whose decay frequency is $-\lambda_j/(2\pi)$, whereas a complex-conjugate pair gives a damped oscillation. Contributions from independent eigenmodes sharing the same eigenvalue are combined in $a_{F,m_F}^{(j)}$ because they have the same time dependence.

Among the Liouvillian modes contributing to the calculated response, the largest total squared amplitude $\sum_{F=3}^{4}\sum_{m_F=-F}^{F}|a_{F,m_F}^{(j)}|^2$ occurs for a twofold-degenerate, nonoscillatory mode family with decay frequency \SI{112.7}{\kilo\hertz}. For this family, we omit the superscript $(j)$ and write its 16 amplitudes as $a_{F,m_F}$. Because the probe is resonant with the $F=4\rightarrow F'=5$ branch, we test whether this mode depends linearly on the initial $m_F$ within $F=4$. The comparison pattern in the 16-initial-state space has entries $m_F$ for $F=4$ and zero for $F=3$; its nonzero entries are $(-4,-3,\ldots,4)$. The fraction of the complete 16-state squared modal amplitude projected onto this pattern is

\begin{equation}
\begin{aligned}
\eta_4
&\equiv
\frac{
\left|
\displaystyle\sum_{m_F=-4}^{4}
m_F a_{4,m_F}
\right|^2
}{
60
\displaystyle\sum_{F=3}^{4}
\sum_{m_F=-F}^{F}
\left|a_{F,m_F}\right|^2
}
=0.983.
\end{aligned}
\label{eq:local_mf_projection}
\end{equation}

Here $60=\sum_{m_F=-4}^{4}m_F^2$ normalizes the $F=4$ comparison pattern, whereas the second sum in the denominator normalizes the complete response over all 16 initial states. Thus $\eta_4=0.983$ means that 98.3\% of the total squared modal amplitude is the $F=4$ component proportional to $m_F$. The far-off-resonant $F=3$ initial states contribute only $2.24\times10^{-10}$ of the total squared amplitude; the remaining 1.7\% lies almost entirely in $F=4$ components that are nonlinear and odd in $m_F$.

\section{Computational implementation}
\label{app:computational_implementation}

The stochastic Liouville--transport solver is implemented in JAX~\cite{JAX2018} and executed through a CUDA-enabled GPU backend. Each calculation propagates all 48 Cs D$_2$ hyperfine--Zeeman sublevels---16 ground and 32 excited states---for $N=64$ atom samples in each of $M=100$ independent realizations. The interval $0\leq\widetilde t\leq10^5$ with $\delta\widetilde t=10^{-3}$ contains $10^8$ sequential time-step intervals per realization. Across all $MN=6400$ atom samples, a production calculation therefore performs $MN\times10^8=6.4\times10^{11}$ updates of 48-component state vectors. The four-stage Runge--Kutta optical update requires $4MN\times10^8=2.56\times10^{12}$ coupling-Hamiltonian actions on such vectors, corresponding to $48\times2.56\times10^{12}=1.23\times10^{14}$ complex output-component evaluations, in addition to diagonal propagation, quantum-jump sampling, ballistic motion, and boundary renewal. JAX/XLA compiles the sequential MCWF kernel, while the atom samples and independent realizations are vectorized across the GPU.

\end{document}